\documentclass[11pt,a4paper,english,showpacs,superscriptaddress]{revtex4}
\usepackage[T1]{fontenc}
\usepackage{amsmath}
\usepackage{graphicx}
\usepackage{amssymb}



\usepackage{babel}
\usepackage{hyperref}

\makeatother

\begin{document}

\title{The $1/N$ Expansion in Noncommutative Quantum Mechanics}

\author{A. F. Ferrari}
\email{alysson.ferrari@ufabc.edu.br}
\affiliation{Universidade Federal do ABC, Centro de Ci\^encias Naturais e Humanas,
Rua Santa Ad\'elia, 166, 09210-170, Santo Andr\'e, SP, Brasil}

\author{M. Gomes}
\email{mgomes@fma.if.usp.br}
\affiliation{Instituto de F\'\i sica, Universidade de S\~ao Paulo, Caixa Postal
66318, 05315-970, S\~ao Paulo - SP, Brazil}

\author{C. A. Stechhahn}
\email{carlos@fma.if.usp.br}
\affiliation{Instituto de F\'\i sica, Universidade de S\~ao Paulo, Caixa Postal
66318, 05315-970, S\~ao Paulo - SP, Brazil}


\begin{abstract}
We study the $1/N$ expansion in noncommutative quantum mechanics
for the anharmonic and Coulombian potentials. The expansion for the
anharmonic oscillator presented good convergence properties, but for
the Coulombian potential, we found a divergent large $N$ expansion
when using the usual noncommutative generalization of the potential.
We proposed a modified version of the noncommutative Coulombian potential
which provides a well-behaved $1/N$ expansion.
\end{abstract}

\maketitle

\section{Introduction}

The so called $1/N$ expansion was proposed in 1974\,\cite{tHooft}
as a scheme to nonperturbatively study QCD in the strong coupling
region. In that context, calculations using a large $N$ expansion
for the group $SU(N)$ have been shown to provide results with good
agreement with experimental data for the $SU(3)$ chromodynamics (see\,\cite{manohar}
for a review). Since then, the $1/N$-expansion was used as an approximation
method which generally gives very accurate results and can be applied
to different fields, including atomic and particle physics. For massless
two dimensional models where severe $IR$ divergences prevent the
use of perturbation methods, the $1/N$-expansion allows to uncover
very interesting peculiarities as dynamical mass generation, dynamical
generation of gauge bosons, confinement and so on\,\cite{coleman}.
Other applications include studies of Bose-Einstein condensation,
stochastic quantization, and noncommutative quantum field theories\,\cite{BoseEinsteinCondensate,cpn,Foerster,grossneveu,MGomes,MGomes2},
to name a few. Recently, the large $N$ limit has become fundamental
in the study of the Maldacena conjecture\,\cite{Maldacena}, which
allows one to obtain nonperturbative information on conformaly invariant
quantum field theories.

The $1/N$-expansion has also been used in Quantum Mechanics for a
large class of potentials, since it produces good results for the
determination of the ground and low excited states energies. The $1/N$-expansion
can be used even when the Hamiltonian cannot be separated in a solvable
part plus a small perturbation; besides that fact, finding energies
and wave functions is achieved by solving iterated algebraic equations,
instead of solving a differential equation. This iterated procedure
can be neatly implemented in any CAS (Computer Algebra System) for
example.

In this paper, we are interested in the application of the $1/N$
expansion in the context of noncommutative quantum mechanical models.
There has been a lot of interest in the last decades in studying theories
defined over a spacetime where coordinates do not commute, in part
following the discovery of noncommutative gauge theories as a low
energy limit of the string theory in certain backgrounds\,\cite{SW}.
The general motivation for spacetime noncommutativity is the idea
that, in distances of the order of the Planck length, the measurement
of the coordinates loses all its sense due to the production of intense
gravitational fields. For this reason, the usual concept of a point
can not be adopted and this suggests the use of position operators
that do not commute\,\cite{Doplicher}.

These motivations rendered to noncommutative spaces a wide variety
of theoretical applications. Several works studying the effects of
the noncommutativity of space in quantum mechanics have appeared recently,
either in nonrelativistic or relativistic situations, see for example\,\cite{gamboa,chaichian,Lubo,shapos,Girotti,Muthukumar,Stechhahn,kup}.
We extend these studies by the use of the $1/N$ expansion applied
to some quantum mechanical potentials. 

This work is organized as follows. In Sec.\,\ref{sec:II} we review
the machinery of the large $N$ expansion in quantum mechanics, and
show how it can be implemented in the noncommutative context. We start
the application of this method in Sec.\,\ref{sec:III}, by studying
the anharmonic oscillator. In Sec\,\ref{sec:IV}, we show that the
$1/N$ expansion diverges when applied to the usual noncommutative
generalization of the Coulombian potential. We argue that this divergence
is due to a strong singularity of the potential at the origin, and
we propose a modification that produces physically significant results.
Our conclusions are summarized in Sec.\,\ref{sec:Conclusions}.

\section{\label{sec:II}Noncommutative Quantum Mechanics and $1/N$ Expansion}

Noncommutative spaces are characterized by the position operators
$\hat{x}_{\mu}$ satisfying the relation\begin{equation}
[\hat{x}_{\mu},\hat{x}_{\nu}]=i\theta_{\mu\nu},\label{eq:01}\end{equation}

\noindent where $\theta_{\mu\nu}$ is a constant antisymmetric matrix
of dimension length squared. Quantum field theories can be formulated
on these spaces, involving field operators which are functions of
$\hat{x}_{\mu}$. However, it is more usual to employ the Weyl's correspondence,
which results in models defined in a commutative spacetime, but with
the pointwise product of fields replaced by the noncommutative Moyal
product \begin{equation}
\phi_{1}(x)*\phi_{2}(x)=\lim_{y\rightarrow x}e^{\frac{i}{2}\theta^{\mu\nu}\frac{\partial}{\partial y^{\mu}}\frac{\partial}{\partial x^{\mu}}}\phi_{1}(y)\phi_{2}(x),\label{eq:02}\end{equation}

\noindent where $\phi_{1}$ and $\phi_{2}$ are two arbitrary functions.

In noncommutative quantum mechanics, a similar approach can be implemented,
so we could study a noncommutative Schrödinger equation involving
a Moyal product $V\left(x\right)*\psi\left(x\right)$, but it is preferred
to perform the change of variables\begin{subequations}\label{eq:03}\begin{align}
\hat{x}_{j} & =x_{j}-\frac{1}{2}\sum_{k}\theta_{jk}p_{k}\\
\hat{p}_{j} & =p_{j}\end{align}
\end{subequations}

\noindent from noncommutative (hatted) operators to new variables
$x_{j}$ and $p_{j}$ satisfying the Heisenberg algebra\begin{subequations}\label{eq:04}\begin{align}
\left[x_{i},x_{j}\right] & =\left[p_{i},p_{j}\right]=0\,,\\
\left[x_{i},p_{j}\right] & =i\hbar\delta_{ij}\,.\end{align}
\end{subequations}In this way, the noncommutative Schrödinger equation
has the standard form, involving the modified potential\begin{equation}
V\left(x_{j}-\frac{1}{2}\sum_{k}\theta^{ij}p_{j}\right).\label{eq:05}\end{equation}
For simplicity, we shall consider a particular form of the $\theta^{ij}$
matrix, where the noncommutativity is nonvanishing only in a particular
spatial plane.

To fix our notations, we shall briefly review the construction of
the $1/N$ expansion for the $N$-dimensional Schrödinger equation\,\cite{Shatz}\begin{equation}
\left[-\frac{1}{2}\nabla_{N}^{2}+V(r)\right]\psi\left(r,\Omega\right)=E\psi\left(r,\Omega\right)\,.\label{eq:06}\end{equation}

\noindent Here, $r^{2}=\sum_{i=1}^{N}x_{i}^{2}$, $\Omega$ is the
set of $N-1$ angular variables, and we are using natural units, so
that $\hbar=c=1$; also, as we use an unitary mass, the energy will
have the unusual dimension of $[L]^{-2}$. The Laplace operator is
given by\begin{equation}
\nabla_{N}^{2}=\frac{\partial^{2}}{\partial r^{2}}+\frac{N-1}{r}\frac{\partial}{\partial r}-\frac{1}{r^{2}}\hat{\Lambda}^{2}\left(N\right)=\Delta_{r}-\frac{1}{r^{2}}\hat{\Lambda}^{2}\left(N\right)\,,\label{eq:07}\end{equation}

\noindent where $\hat{\Lambda}^{2}(N)$ is the generalized angular
momentum squared (we define $\hat{\Lambda}^{2}(1)=0$). The wavefunction
of this system can be separated in radial and angular parts,\begin{equation}
\psi\left(r,\Omega\right)=R_{n\ell}\left(r\right)Y\left(\Omega\right)\,,\label{eq:08}\end{equation}

\noindent where $R_{n\ell}(r)$ is labeled by two quantum numbers
$n$ and $\ell$, and the generalized spherical harmonics $Y\left(\Omega\right)=Y_{\ell_{1},\ell_{2},\ldots,\ell_{N-2},\ell_{N-1}}\left(\phi_{1},\phi_{2},\ldots,\phi_{N-1}\right)$
are labeled by $N-1$ quantum numbers $\ell_{1},\ell_{2},\ldots,\ell_{N-2},\ell_{N-1}=\ell$.
Replacing Eqs.\,(\ref{eq:07}) and\,(\ref{eq:08}) in Eq.\,(\ref{eq:06}),
the Schrödinger equation is separated into a radial\begin{equation}
\left[-\frac{1}{2}\left(\frac{d^{2}}{dr^{2}}+\frac{N-1}{r}\frac{d}{dr}\right)+\frac{\ell(\ell+N-2)}{2r^{2}}+V(r)\right]R_{n\ell}(r)=ER_{n\ell}(r)\label{eq:09}\end{equation}

\noindent and an angular equation\begin{equation}
\hat{\Lambda}^{2}(N)Y(\Omega)=\ell(\ell+N-2)Y(\Omega),\label{eq:10}\end{equation}

\noindent where the allowed quantum numbers for the generalized spherical
harmonics are $\ell=0,1,2,\ldots$, $\ell_{j}=0,1,2,\ldots,\ell_{j+1}$
for $j=2,3,\ldots,N-2$, and $\ell_{1}=m=-\ell_{2},-\ell_{2}+1,\ldots,\ell_{2}-1,l_{2}$\,\cite{louck}.
The label $\ell_{1}$ corresponds to the eigenvalue of the $L_{12}$
component of the angular momentum, and shall be further called $m$
for similarity with the three-dimensional case.

It is customary to eliminate the first order derivative in Eq.\,(\ref{eq:09})
by means of the substitution\begin{equation}
\eta\left(r\right)=r^{\frac{N-1}{2}}R_{n\ell}\left(r\right),\label{eq:11}\end{equation}

\noindent which leads to\begin{equation}
\left\{ -\frac{1}{2}\frac{d^{2}}{dr^{2}}+k^{2}\left[\frac{(1-1/k)(1-3/k)}{8r^{2}}+\hat{V}(r)\right]\right\} \eta\left(r\right)=E\,\eta\left(r\right)\,,\label{eq:12}\end{equation}

\noindent where $k=N+2\ell$, and the normalized potential $\hat{V}(r)$
is defined as $\hat{V}=V/k^{2}$. 

The above equation shows that $k^{2}$ behaves as a mass and the kinetic
term can be disregarded in the $k\rightarrow\infty$ limit. Thus,
when $k$ is very large, the ground state of the system is located
at the minimum, $r_{0}$, of the effective potential\begin{equation}
V_{\mbox{eff}}\left(r\right)=\frac{1}{8r^{2}}+\hat{V}\left(r\right),\label{eq:15}\end{equation}
so that the ground state energy is, in the leading approximation,\begin{equation}
E_{0}=k^{2}V_{\mbox{eff}}\left(r_{0}\right)\,,\label{eq:14}\end{equation}

\noindent with $r_{0}$ defined by\begin{equation}
\left.\frac{dV_{\mbox{eff}}}{dr}\right|_{r=r_{0}}=0\,.\label{eq:16}\end{equation}

To obtain higher-order corrections to the ground-state energy, it
is convenient to redefine the radial wave function and to rescale
the radial coordinate according to\begin{equation}
\eta\left(r\right)=\exp A\left(r\right)\,,\label{eq:17}\end{equation}
and

\noindent \textbf{\begin{equation}
u=\frac{r}{r_{0}}\,,\label{eq:18}\end{equation}
}thus obtaining a Riccati equation\begin{equation}
-\frac{1}{2r_{0}^{2}}\left[U^{2}\left(u\right)+U^{\prime}\left(u\right)\right]+k^{2}V_{\mbox{eff}}(u)+\left(-\frac{k}{2}+\frac{3}{8}\right)\frac{1}{r_{0}^{2}u^{2}}=E,\label{eq:19}\end{equation}

\noindent where $U=\frac{1}{r_{0}}\frac{dA}{du}$ and $U^{\prime}=\frac{1}{r_{0}}\frac{dU}{du}$.
This equation can be solved in a power series in $1/k$, using\begin{subequations}\label{eq:20}\begin{align}
E & =\sum_{n=-2}^{\infty}E^{(n)}k^{-n}\,,\\
U & =\sum_{n=-1}^{\infty}U^{(n)}k^{-n}\,.\end{align}
\end{subequations}Replacing these expressions in Eq.\,(\ref{eq:19})
and equating to zero the coefficient of each power in $1/k$, we get
the following set of equations, \begin{equation}
-\frac{1}{2r_{0}^{2}}U^{(-1)}(u)U^{(-1)}(u)+V_{\mbox{eff}}(u)=E^{(-2)}\,,\label{eq:21}\end{equation}

\begin{equation}
-U^{(-1)}(u)U^{(0)}(u)=r_{0}^{2}E^{(-1)}+\frac{1}{2u^{2}}+\frac{1}{2}U^{(-1)\prime}(u)\,,\label{eq:22}\end{equation}

\begin{equation}
-U^{(-1)}(u)U^{(1)}(u)=r_{0}^{2}E^{(0)}-\frac{3}{8u^{2}}+\frac{1}{2}\left[U^{(0)\prime}(u)+U^{(0)}(u)U^{(0)}(u)\right]\,,\label{eq:23}\end{equation}

\begin{equation}
-U^{(-1)}(u)U^{(n+1)}(u)=r_{0}^{2}E^{(n)}+\frac{1}{2}\left[U^{(n)\prime}(u)+\sum_{m=0}^{n}U^{(m)}(u)U^{(n-m)}(u)\right]\,,\quad n>0,\label{eq:24}\end{equation}

\noindent which, in principle, can be solved iteratively up to any
order in $1/k$. By evaluating Eq.\,(\ref{eq:21}) at $u=1$ we reobtain
Eq.\,(\ref{eq:14}) for the leading approximation to the ground state
energy,\begin{equation}
E^{(-2)}=V_{\mbox{eff}}\left(r_{0}\right)\,.\label{eq:25}\end{equation}
By replacing this result back in Eq.\,$(\ref{eq:21})$ and solving
for $U^{(-1)}$, one gets\begin{equation}
U^{(-1)}(u)=-\sqrt{2r_{0}^{2}\left(V_{eff}(u)-E^{(-2)}\right)}.\label{eq:26}\end{equation}

\noindent In this equation the $(-)$ signal has to be chosen so that
the function $U=dA/dr$ is positive for $u<1$ and negative for $u>1$,
since the wave function has a maximum at the point $r_{0}$. This
procedure should be repeated order by order in $k$ to obtain the
higher order corrections.

Two remarks are now in order: first, excited states are considered
by modifying the ansatz in Eq.\,(\ref{eq:17}) to\begin{equation}
\eta_{n}\left(r\right)=\left[\prod_{j=1}^{n}\left(r-r_{j}\right)\right]\exp A\left(r\right)\,,\label{eq:27}\end{equation}
to account for the $n$ nodes of the $n$th excited state\,\cite{Shatz}.
Second, one alternative way to obtain an $1/N$ expansion would be
to expand Eq.\,(\ref{eq:12}) in a power series in \begin{equation}
x=k^{p}\frac{r-r_{0}}{r_{0}}\,,\label{eq:28}\end{equation}
where $p$ is some (positive) constant\,\cite{kalara}. For $k$
very large, Eq.\,(\ref{eq:28}) shows that the wave-function should
be highly concentrated around $x=0$, so it could be calculated as
a power series around the minimum of the potential. At the dominant
order, Eq.\,(\ref{eq:12}) reduces to an harmonic oscillator equation.
Higher order corrections in $k$ are included using standard perturbation
theory. In this work, we shall use the approach based on Eq.\,(\ref{eq:21}),
where the eigenvalues and eigenfunctions are expanded in powers of
$1/k$ and solved order by order by algebraic procedures, and no further
approximation schemes are necessary. This scheme can be extended to
the noncommutative situation by using the modified potential specified
in Eq.\,(\ref{eq:05}).

\section{\label{sec:III}The Noncommutative Anharmonic Oscillator}

As a first example, we shall now consider a noncommutative anharmonic
oscillator in $N$-dimensional space, with an Hamiltonian defined
as\,\cite{Koudinov} \begin{equation}
\mathcal{\hat{H}}=\sum_{i=1}^{N}\left(\frac{1}{2}\hat{p}_{i}^{2}+\frac{\omega_{0}^{2}}{2}\hat{x}_{i}^{2}+\frac{g}{N}\left(\hat{x}_{i}^{2}\right)^{2}\right)\,.\label{eq:29}\end{equation}

\noindent After performing the change of variables of Eq.\,(\ref{eq:03}),
we obtain the following Schrödinger equation,\begin{equation}
\left[-\frac{1}{2}\left(\nabla_{N}^{2}-\frac{1}{r^{2}}\hat{\Lambda}^{2}\left(N\right)\right)+V\left(x,p\right)\right]R_{n\ell}\left(r\right)Y\left(\Omega_{N}\right)=\mathcal{E}\, R_{n\ell}\left(r\right)Y\left(\Omega_{N}\right)\,,\label{eq:30}\end{equation}

\noindent with the potential\begin{equation}
V\left(x,p\right)=\frac{\omega_{0}^{2}}{2}\left(r^{2}-\sum_{ij}\theta_{ij}x_{i}p_{j}\right)+\frac{g}{N}\left(r^{4}-2r^{2}\sum_{i,j}\theta_{ij}x_{i}p_{j}\right)\,.\label{eq:31}\end{equation}

The potential in Eq.\,(\ref{eq:31}) has two parameters, $\omega_{0}$
and $g^{1/3}$, with dimensions of energy. Following\,\cite{Koudinov},
we introduce a new parameter $\omega$ that will fix the energy scale,
and we shall work with the adimensional energy $E=\mathcal{E}/\omega$
and adimensional coupling constant $\lambda=g/\omega^{3}$. The relation
between $g$, $\omega_{0}$, $\lambda$ and $\omega$ is given by\begin{subequations}\label{eq:32}\begin{align}
\frac{\omega_{0}^{2}}{\omega^{2}} & =1-2\lambda\,,\\
\lambda & =\frac{g}{\omega^{3}}\,.\end{align}
\end{subequations}Since we are not interested in studying symmetry
breaking we shall consider $\omega_{0}^{2}>0$, thus $\lambda$ is
constrained by $0\le\lambda\le1/2$. We shall perform the rescaling
$x_{i}\rightarrow x_{i}/\sqrt{\omega}$ and $\theta\rightarrow\theta/\omega$
to obtain the Schrödinger equation in the very same form as Eq.\,(\ref{eq:30}),
but involving the adimensional energy $E$ in the right hand side,
and the potential\begin{equation}
V\left(x,p\right)=\frac{1-2\lambda}{2}\left(r^{2}-\sum_{ij}\theta_{ij}x_{i}p_{j}\right)+\frac{\lambda}{N}\left(r^{4}-2r^{2}\sum_{i,j}\theta_{ij}x_{i}p_{j}\right)\,,\label{eq:33}\end{equation}
in the left hand side.

For simplicity, we assume that the only nonvanishing component of
$\theta_{ij}$ is $\theta_{12}=-\theta_{21}=\theta$. In this case,
$\sum_{ij}\theta_{ij}x_{i}p_{j}$ simplifies to $\theta L_{12}$,
the component of the angular momentum perpendicular to the plane of
noncommutative coordinates.

As discussed in Sec.\,\ref{sec:II}, the first label $\ell_{1}=m$
of the generalized spherical harmonic $Y(\Omega)$ corresponds to
the eigenvalue of $L_{12}$. By using this fact in Eq.\,(\ref{eq:30})\textbf{,}
we can finally write the potential for the noncommutative anharmonic
oscillator as\begin{align}
V\left(r\right)= & \left(\frac{1-2\lambda}{2}-\frac{2\theta m\lambda}{N}\right)r^{2}+\frac{\lambda}{N}r^{4},\label{eq:34}\end{align}
apart from a constant term $-\theta m\left(1-2\lambda\right)/2$.
The effective potential reads \begin{equation}
V_{\mbox{eff}}(r)=\frac{1}{8r^{2}}+\frac{1}{k^{2}}\left[\left(\frac{1-2\lambda}{2}-\frac{2\theta m\lambda}{N}\right)r^{2}+\frac{\lambda}{N}r^{4}\right]\,,\label{eq:35}\end{equation}
and its minimum is located at $r_{0}$ satisfying the equation,\begin{equation}
-\frac{1}{4}+(1-2\lambda)\frac{r_{0}^{4}}{k^{2}}+\frac{4\lambda}{k^{2}N}r_{0}^{6}-\frac{4\theta m\lambda}{N}\frac{r_{0}^{4}}{k^{2}}=0\,.\label{eq:36}\end{equation}
Since the above equation involves both $N$ and $k=N+2\ell$, we choose
to find its solution as a power series in $1/N$, by defining $r_{0}=\sqrt{\frac{N}{2}}\bar{r}_{0}$,
\begin{equation}
\bar{r}_{0}=1+r_{1}\left(\frac{1}{N}\right)+r_{2}\left(\frac{1}{N}\right)^{2}+\ldots\,,\label{eq:38}\end{equation}

\noindent where $\sqrt{N/2}$ is the solution for the commutative,
$\ell=0$ case. We quote here the explicit form for $r_{0}$ up to
the first order in $1/N$ and $\theta$, \begin{equation}
r_{0}=\sqrt{\frac{N}{2}}\left[1+\left(\frac{\ell}{1+\lambda}+\frac{\theta\lambda m}{1+\lambda}\right)\frac{1}{N}+\cdots\right]\,.\label{eq:39}\end{equation}

The Riccati equation for the noncommutative anharmonic oscillator,
in terms of the variable $u=r/\bar{r}_{0}$, reads\begin{equation}
-\frac{1}{N\overline{r}_{0}^{2}}\left(U^{\prime}+U^{2}\right)+NW(u)+\left(\frac{3}{4}\frac{1}{N\overline{r}_{0}^{2}}-\frac{1+\epsilon}{\overline{r}_{0}^{2}}\right)\frac{1}{u^{2}}=E\,,\label{eq:40}\end{equation}

\noindent where\begin{equation}
W(u)=\frac{1}{4}\left(\frac{1+\epsilon}{\bar{r}_{0}}\right)^{2}\frac{1}{u^{2}}+\left[\frac{1-2\lambda}{2}-\frac{\theta\lambda m\epsilon}{\ell}\right]\frac{\bar{r}_{0}^{2}u^{2}}{2}+\frac{\lambda}{4}\bar{r}_{0}^{4}u^{4}\,.\label{eq:41}\end{equation}
The leading order contribution to the ground state energy in the noncommutative
case coincides with the commutative one,\begin{equation}
E^{\left(-2\right)}=NW(1)=N\left(\frac{2-\lambda}{4}\right),\label{eq:42}\end{equation}
since the $\theta$-dependent term in Eq.\,(\ref{eq:35}) is subleading
in the $1/N$ expansion. By subtracting $E^{\left(-2\right)}$ from
both sides of the Eq.\,(\ref{eq:40}), we obtain\begin{align}
-\frac{1}{N\overline{r}_{0}^{2}}\left(U^{\prime}+U^{2}\right)+N\left[W\left(u\right)-\left(\frac{2-\lambda}{4}\right)\right]\nonumber \\
+\left[\left(\frac{3}{4}-2\ell\right)\frac{1}{N}-1\right]\frac{1}{\bar{r}_{0}^{2}u^{2}} & =E-N\left(\frac{2-\lambda}{4}\right)=E^{\prime}\,.\label{eq:43}\end{align}
Each term involving $\overline{r}_{0}$ is expanded as a power series
in $1/N$, and so will be the energy $E^{\prime}$\begin{equation}
E^{\prime}=E^{\left(0\right)}+\sum_{j\geq1}E^{\left(j\right)}\frac{1}{N^{j}}\label{eq:44}\end{equation}
and \begin{equation}
U\left(u\right)=NU^{(-1)}\left(u\right)+U^{(0)}\left(u\right)+\sum_{j\geq1}U^{(j)}\left(u\right)\frac{1}{N^{j}}\,.\label{eq:45}\end{equation}

With these formulae, one is able to calculate $E^{\left(j\right)}$
and $U^{\left(j\right)}$ to an arbitrary $1/N$ order, at least in
principle. In practice, calculations by hand are amenable up to order
$1/N$, further corrections can be calculated using a computer. For
example, the first two corrections to the ground state energy are

\noindent \begin{align}
E^{\prime}= & -1+\sqrt{1+\lambda}+\ell-\theta\lambda m\nonumber \\
 & +\left[\frac{\lambda\left(4\ell^{2}\lambda+4\ell^{2}-8\ell\lambda+12\ell\sqrt{\lambda+1}-8\ell+\lambda-12\sqrt{\lambda+1}+12\right)}{4(\lambda+1)^{2}}\right.\nonumber \\
 & +\left.\frac{\lambda\left(-2m\ell\lambda-2m\ell+2m\lambda+m\lambda\sqrt{\lambda+1}-2m\sqrt{\lambda+1}+2m\right)}{(\lambda+1)^{2}}\theta\right]\frac{1}{N}\,.\label{eq:46}\end{align}
Equation\,(\ref{eq:46}) correctly reproduces the results of the
commutative anharmonic oscillator when $\ell=0$ and $\theta=0$\,\cite{Koudinov}. 

\begin{figure}[p]
\begin{centering}
\includegraphics[width=0.8\columnwidth]{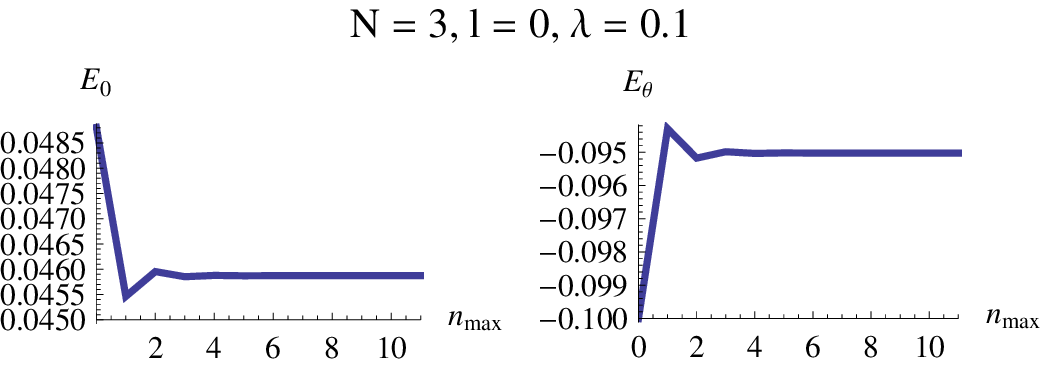}
\par\end{centering}

\begin{centering}
\includegraphics[width=0.8\columnwidth]{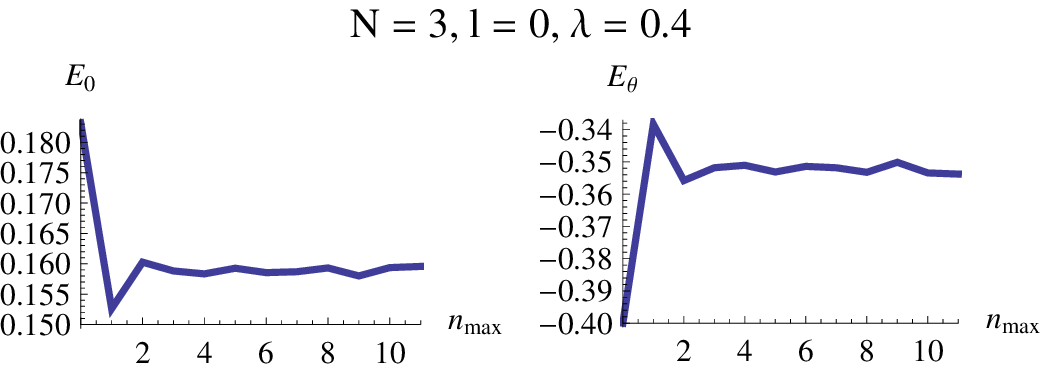}
\par\end{centering}

\begin{centering}
\includegraphics[width=0.8\columnwidth]{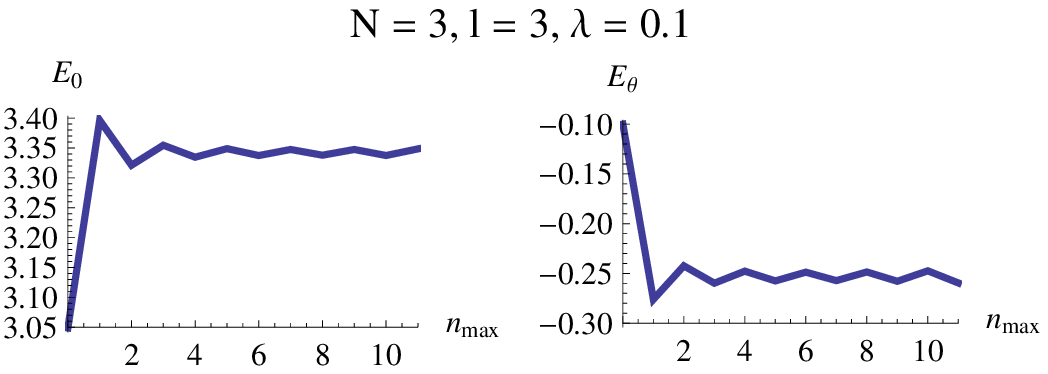}
\par\end{centering}

\caption{\label{fig:1}Ground-state (adimensional) energy for the noncommutative
anharmonic potential calculated up to order $n_{\mbox{max}}$, in
the form $E^{\prime}=E_{0}+E_{\theta}\theta$, for different values
of $\ell$ and $\lambda$.}

\end{figure}

We found that, even using a standard desktop computer, the fully analytical
calculation could not be done beyond the $1/N^{3}$ order in reasonable
time; however, by choosing some particular numerical values for $\ell$
and $\lambda$, one can quickly calculate the corrections up to $1/N^{12}$
or even more. Some results are shown in graphical form in Fig.\,(\ref{fig:1}):
the horizontal axis is the order of the $1/N$ expansion used, i.e.,
for each $n_{\mbox{max}}$ we calculate the adimensional energy $E^{\prime}=E_{0}+\theta E_{\theta}$
up to order $1/N^{n_{\mbox{max}}}$. These graphs suggests that the
convergence is quite good, at least for small enough $\lambda$ and
for $\ell=0$. For higher $\ell$, the results are not so stable,
and the reason is clear from Eq.\,(\ref{eq:39}): the first correction
for $r_{0}$ is of order $\ell/N$, so Eq.\,(\ref{eq:38}) does not
provide a good approximation to $r_{0}$ if $\ell$ is not much smaller
than $N$.

\section{\label{sec:IV}The Noncommutative Coulombian Potential}

We now focus on the noncommutative generalization of the Coulombian
potential, which is usually given in terms of the noncommutative coordinates
$\hat{x}$ as \begin{equation}
V\left(\hat{x}\right)=-\frac{Ze^{2}}{\sqrt{\hat{x}\hat{x}}}\,.\label{eq:47}\end{equation}

\noindent As described in Sec.\,\ref{sec:II}, the customary way
to work with this potential is by means of the change of variables
in Eq.\,(\ref{eq:03}), which yields\begin{equation}
V\left(x,p\right)=-\frac{Ze^{2}}{\sqrt{r^{2}-\sum_{ij}\theta_{ij}x_{i}p_{j}+\frac{1}{4}\theta_{j\ell}\theta_{jk}p_{\ell}p_{k}}}\,.\label{eq:48}\end{equation}
A direct treatment of this potential is quite difficult from a technical
viewpoint. This is why, in the literature\,\cite{chaichian,kao,adorno},
it has been studied using standard perturbation theory after an expansion
up to the first order in $\theta$ as follows,\begin{equation}
V=-\frac{Ze^{2}}{r}\left[1+\frac{1}{2r^{2}}\sum_{ij}\theta_{ij}x_{i}p_{j}\right]\,.\label{eq:49}\end{equation}
We notice that the noncommutative correction to the potential behaves
as $1/r^{3}$, so it is more singular at the origin than the one in
Eq.\,(\ref{eq:47}). We shall also stress that Eq.\,(\ref{eq:49})
is not a valid approximation when $r$ is very small. Such issue has
not been considered in the literature so far because in standard perturbation
theory one is interested in integrals of the general form $\left\langle \psi|V\left(r\right)|\psi\right\rangle $,
which are actually regular despite the singularity at the origin.
As we shall see, when using the $1/N$ expansion, this singular behavior
near the origin will be a major issue we will have to deal with. In
this work, we will show how to generalize the potential in Eq.\,(\ref{eq:49})
so that it produces a meaningful $1/N$ expansion. 

Hereafter, all our expressions are calculated up to the first order
in $\theta$. As before, we shall consider the particular case $\theta_{12}=\theta$
with all other components of the matrix $\theta_{ij}$ vanishing,
such that $\sum_{ij}\theta_{ij}x_{i}p_{j}=\theta L_{12}$. In this
case, $V\left(x,p\right)$ reduces to \begin{equation}
V\left(r\right)=-\frac{Ze^{2}}{r}\left[1+\theta\frac{L_{12}}{2r^{2}}\right]\,.\label{eq:50}\end{equation}

We start by taking the potential in Eq.\,(\ref{eq:50}) as our starting
point. By means of the change of variables\begin{equation}
\rho=4Z\hat{e}^{2}\, r\quad;\quad\hat{\theta}=\theta\left(4Z\hat{e}^{2}\right)^{2},\label{eq:53}\end{equation}

\noindent the Schrödinger equation becomes\begin{equation}
\left[-\frac{1}{2}\frac{d^{2}}{d\rho^{2}}+k^{2}V_{\mbox{eff}}\left(\rho\right)+\left(\frac{3}{8}-\frac{1}{2}k\right)\frac{1}{\rho^{2}}\right]\eta\left(\rho\right)=E\eta\left(\rho\right)\,,\label{eq:54}\end{equation}

\noindent with the effective potential\begin{equation}
V_{\mbox{eff}}\left(\rho\right)=\frac{1}{8\rho^{2}}-\frac{1}{4\rho}-\frac{\hat{\theta}m}{8\rho^{3}}\,.\label{eq:55}\end{equation}
Here, $\hat{e}^{2}=e^{2}/k^{2}$, $m$ is the eigenvalue of $L_{12}$,
and the adimensional energy $E$ is measured in units of $16Z^{2}\hat{e}^{4}$.
For simplicity of notation, we shall drop the hat in $\hat{\theta}$
from now on.

\begin{figure}[p]
\begin{centering}
\includegraphics[width=0.8\columnwidth]{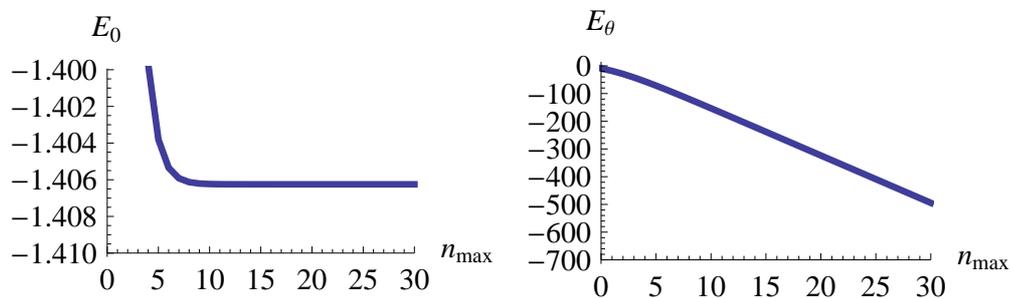}
\par\end{centering}

\caption{\label{fig:2}Ground-state energy of the noncommutative Coulombian
potential calculated up to the order $1/k^{n_{\mbox{max}}}$.}

\end{figure}

The minimum of the effective potential in Eq.\,(\ref{eq:55}) is
located at\begin{equation}
\rho_{0}=1-\frac{3\theta m}{2}\,.\label{eq:56}\end{equation}
The leading approximation to the ground state energy is given by\begin{equation}
E^{\left(-2\right)}=V_{\mbox{eff}}\left(\rho_{0}\right)=-\frac{(1+\theta\, m)}{8}\,.\label{eq:57}\end{equation}
To find higher-order corrections, we solve the Riccati equation \begin{equation}
-\frac{1}{2\rho_{0}^{2}}\left[U^{2}\left(u\right)+U^{\prime}\left(u\right)\right]+k^{2}V_{\mbox{eff}}\left(u\right)+\left(\frac{3}{8}-\frac{1}{2}k\right)\frac{1}{u^{2}}=E\,,\label{eq:58}\end{equation}
where $u=\rho/\rho_{0}$. Both $U$ and $E$ are expanded in orders
of $1/k$ as in Eq.\,(\ref{eq:20}). In the leading order, the wavefunction
is given by\begin{align}
U^{\left(-1\right)}\left(u\right) & =-\sqrt{2\rho_{0}^{2}\left(V_{\mbox{eff}}\left(u\right)-V_{\mbox{eff}}\left(1\right)\right)}\nonumber \\
 & =\frac{1-u}{2u}+\frac{\theta\, m}{4u^{2}}\left(u-1\right)\left(2u+1\right)\,,\label{eq:60}\end{align}
whose derivative at $u=1$, using Eq.\,(\ref{eq:22}), gives the
subleading correction to the energy of the ground state, \begin{equation}
E^{(-1)}=-\frac{1}{4}\left(1+\frac{9}{2}\theta m\right)\,.\label{eq:61}\end{equation}
Inserting back this value of $E^{(-1)}$ in Eq. (\ref{eq:22}), one
obtains the subleading contribution to the wavefunction,\begin{equation}
U^{\left(0\right)}\left(u\right)=-\frac{u+1}{2u}-\frac{5u^{2}+6u+3}{4u^{2}}\theta m\,.\label{eq:62}\end{equation}
For the noncommutative Coulombian potential, this procedure can be
repeated to higher orders in $1/k$. A simple computer program was
used to calculate both $E^{\left(n\right)}$ and $U^{\left(n\right)}$
up to $n_{\mbox{max}}\sim50$ in a few seconds. Up to order $1/k^{10}$,
the energy of the ground state was calculated as\begin{align}
E= & -\frac{\kappa}{4}-\frac{3}{8}-\frac{1}{2\kappa}-\frac{5}{8\kappa^{2}}-\frac{3}{4\kappa^{3}}-\frac{7}{8\kappa^{4}}-\frac{1}{\kappa^{5}}\nonumber \\
 & -\frac{9}{8\kappa^{6}}-\frac{5}{4\kappa^{7}}-\frac{11}{8\kappa^{8}}-\frac{3}{2\kappa^{9}}-\frac{13}{8\kappa^{10}}\nonumber \\
 & +\theta\left(-\frac{9\kappa}{8}-\frac{49}{8}-\frac{211}{8\kappa}-\frac{199}{2\kappa^{2}}-\frac{1385}{4\kappa^{3}}-\frac{4579}{4\kappa^{4}}-\frac{14645}{4\kappa^{5}}\right.\nonumber \\
 & \left.-\frac{91667}{8\kappa^{6}}-\frac{282815}{8\kappa^{7}}-\frac{864359}{8\kappa^{8}}-\frac{2625269}{8\kappa^{9}}-\frac{3970323}{4\kappa^{10}}\right)\,.\label{eq:63}\end{align}
From this result, we see a quite different behavior for the $\theta$-independent
terms and for the $\theta$-dependent ones. This result is graphically
represented in Fig.\,\ref{fig:2}; the $\theta$-independent contribution
to the energy converges quickly and this convergence is very stable
for higher orders of $1/k$, while the $\theta$-dependent ones badly
diverges. 

Divergences of the $1/N$ expansion are not surprising, since the
expansion is usually stable up to some order but it diverges at higher
orders (see for example\,\cite{bohr}). However, in our case, there
is no convergence at all, so the $1/N$ expansion does not provide
a useful calculational scheme. However, it is interesting to notice
that this problem is restricted to the $\theta$-dependent part of
the energy, so its origin is in the noncommutative part of the potential.
The main particularity of this term is the stronger singularity at
the origin, and we will now show that modifying the potential in Eq.\,(\ref{eq:50})
to soften this singularity will indeed avoid the divergence of the
$1/N$ expansion.

We propose a modified version of the noncommutative Coulombian potential
as follows,\begin{equation}
V\left(r\right)=-\frac{Z\hat{e}^{2}}{r}\left[1+\frac{m}{2r^{2}}\left(1-e^{-\alpha r^{\beta}}\right)\frac{\theta}{k}\right]\,.\label{eq:64}\end{equation}
With this modification, the noncommutative part of the potential behaves
as $1/r^{3-\beta}$ near the origin, so it is actually less divergent
than the usual Coulombian potential if $\beta>2$. The factor $\alpha$
has dimension $\left[\mbox{length}\right]^{-\beta}$, so it defines
the characteristic length scale of the modification we are introducing.
In proposing this potential, we have also redefined the noncommutativity
parameter $\theta$ as $\theta\rightarrow\theta/k$: this is needed
because, due to the exponential function in Eq.\,(\ref{eq:64}),
the equation defining $r_{0}$ would be transcendental and no analytic
solution could be found\,\cite{foot1}. With the rescaling $\theta\rightarrow\theta/k$,
the effective potential does not include any $\theta$-dependence,\begin{equation}
V_{\mbox{eff}}\left(r\right)=\frac{1}{8r^{2}}-\frac{Z\hat{e}^{2}}{r}\,,\label{eq:65}\end{equation}
and all the modification due to the noncommutativity enters through
subleadings corrections obtained from the Riccati equation,\begin{equation}
-\frac{1}{2r_{0}^{2}}\left[U^{2}\left(u\right)+U^{\prime}\left(u\right)\right]+k^{2}V_{\mbox{eff}}(u)-\frac{k}{2}\left[\frac{1}{r_{0}^{2}u^{2}}+\frac{Z\hat{e}^{2}}{r_{0}^{3}u^{3}}\theta m\left(1-e^{-\alpha r^{\beta}}\right)\right]+\frac{3}{8}\frac{1}{r_{0}^{2}u^{2}}=E.\label{eq:66}\end{equation}
We remark that such rescalings are usual in $1/k$ expansion, as discussed
in\,\cite{kalara}.

By redefining coordinates as\begin{equation}
\rho=4Z\hat{e}^{2}\, r\hspace{0.3cm};\hspace{0.3cm}\tilde{\theta}=\theta\left(4Z\hat{e}^{2}\right)^{2}\hspace{0.3cm};\hspace{0.3cm}\tilde{\alpha}=\frac{\alpha}{(4Z\hat{e}^{2})^{\beta}}\hspace{0.3cm};\hspace{0.3cm}\tilde{E}=E/\left(4Z\hat{e}^{2}\right)^{2}\label{eq:67}\end{equation}
we rewrite Eq.\,(\ref{eq:66}) as (dropping the tildes)\begin{equation}
-\frac{1}{2}\left[U^{2}\left(\rho\right)+U^{\prime}\left(\rho\right)\right]+k^{2}V_{\mbox{eff}}\left(\rho\right)-k\left[\frac{1}{2\rho^{2}}+\frac{\theta m}{8\rho^{3}}\left(1-e^{-\alpha\rho^{\beta}}\right)\right]+\frac{3}{8\rho^{2}}=E\,,\label{eq:68}\end{equation}
where the effective potential is given by Eq.\,(\ref{eq:65}). The
minimum of $V_{\mbox{eff}}\left(\rho\right)$ is located at $\rho_{0}=1$,
the leading order energy reads $E^{\left(-2\right)}=-1/4$ and\begin{equation}
U^{(-1)}=\frac{1-\rho}{2\rho}\,,\label{eq:69}\end{equation}
which is the same as the commutative case. We follow the procedure
outlined in Sec.\,\ref{sec:II} to calculate higher order corrections
to $E$ and $U$, the only modification is in Eq.\,(\ref{eq:22})
for the subleading order, which is modified to\begin{equation}
-\frac{1}{2}\left[2U^{(-1)}U^{(0)}+U^{(-1)\prime}\right]-\frac{1}{2\rho^{2}}-\frac{\theta m}{8\rho^{3}}\left(1-e^{-\alpha\rho^{\beta}}\right)=E^{(-1)}\,,\label{eq:70}\end{equation}
now including the noncommutative correction, which will therefore
appear starting in the subleading order. 

\begin{figure}[p]
\begin{centering}
\includegraphics[width=0.8\columnwidth]{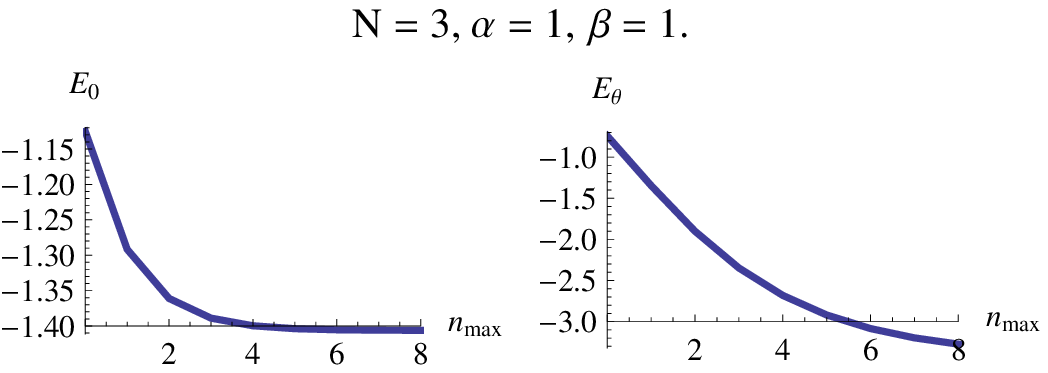}
\par\end{centering}

\begin{centering}
\includegraphics[width=0.8\columnwidth]{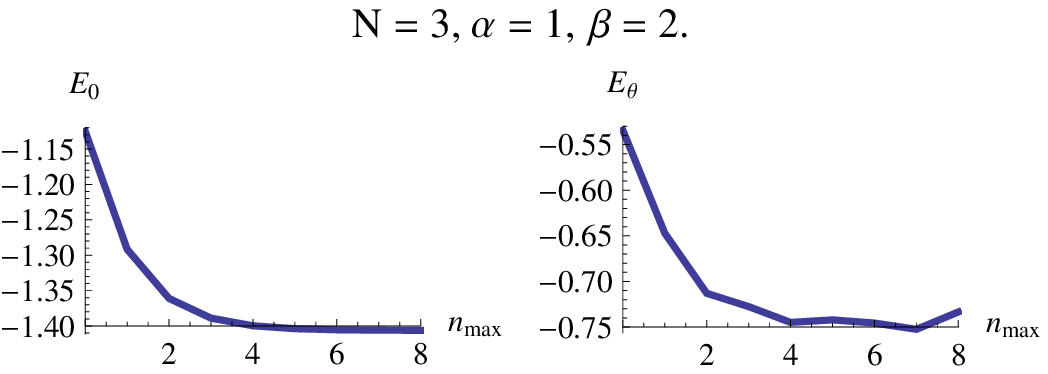}
\par\end{centering}

\begin{centering}
\includegraphics[width=0.8\columnwidth]{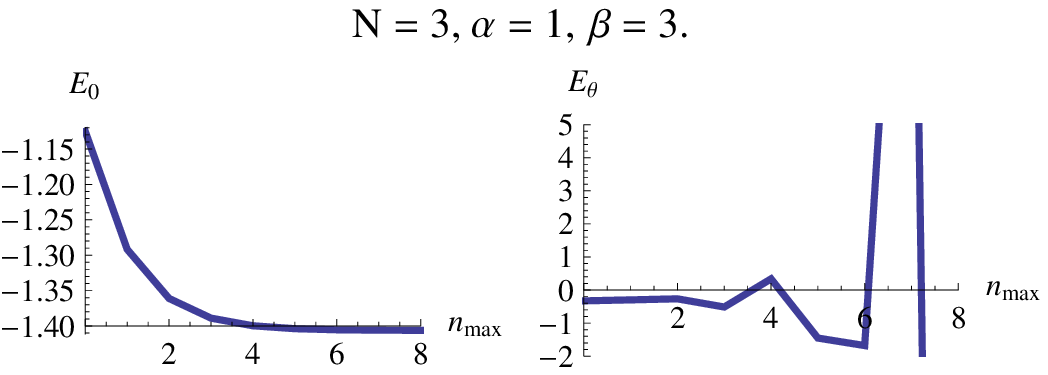}
\par\end{centering}

\caption{\label{fig:3}Ground-state energy of the modified noncommutative Coulombian
potential calculated up to the order $1/k^{n_{\mbox{max}}}$, for
different values of $\alpha$ and $\beta$.}

\end{figure}

\begin{figure}[p]
\begin{centering}
\includegraphics[width=0.8\columnwidth]{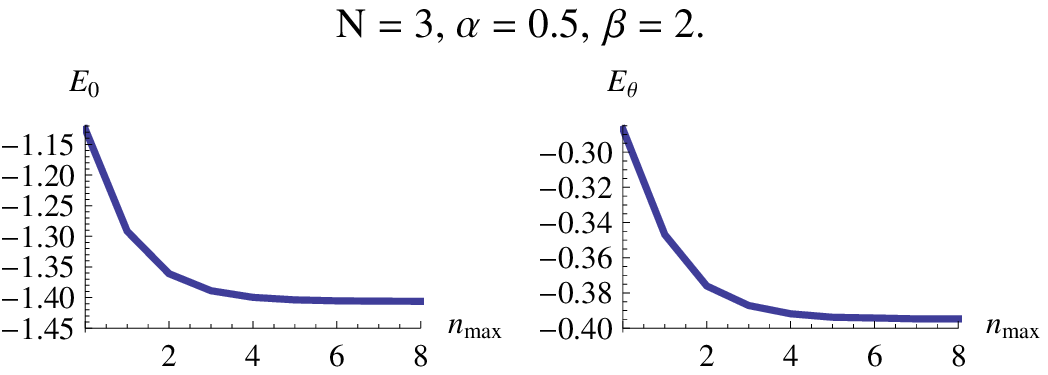}
\par\end{centering}

\begin{centering}
\includegraphics[width=0.8\columnwidth]{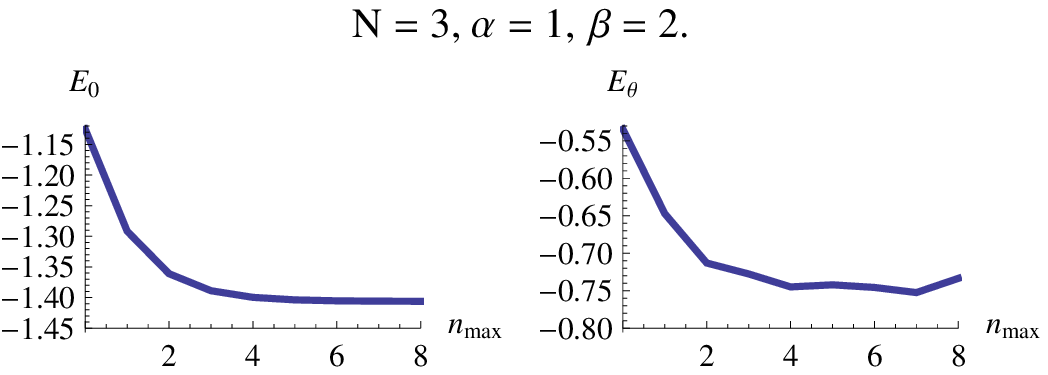}
\par\end{centering}

\begin{centering}
\includegraphics[width=0.8\columnwidth]{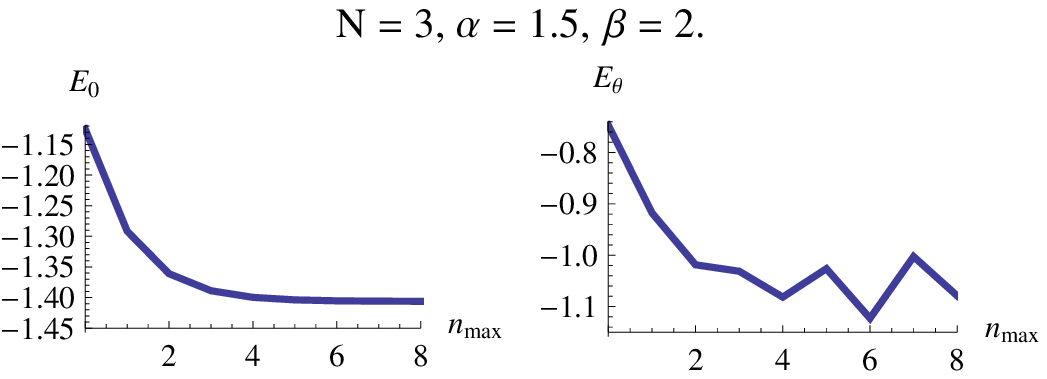}
\par\end{centering}

\caption{\label{fig:4}Same as in Fig.\,\ref{fig:3}, for fixed $\beta=2$
and different values of $\alpha$. }

\end{figure}

We have calculated analytically the energy up to the order $1/k^{8}$
using a \emph{Mathematica} program, and we plotted $E=E_{0}+\theta E_{\theta}$
for several orders of the $1/k$ expansion, looking for values for
$\alpha$ and $\beta$ which would provide a reasonably convergence
and stability of the $1/k$ expansion. Some results are depicted in
Fig.\,\ref{fig:3}. We found that for $\alpha$ of order unity and
$\beta=2$ we obtained the best convergence results. In this situation,
we obtained\begin{align}
E= & -0.25\kappa-0.375-\frac{0.5}{\kappa}-\frac{0.625}{\kappa^{2}}-\frac{0.75}{\kappa^{3}}-\frac{0.875}{\kappa^{4}}-\frac{1.0}{\kappa^{5}}-\frac{1.125}{\kappa^{6}}-\frac{1.25}{\kappa^{7}}-\frac{1.375}{\kappa^{8}}\nonumber \\
 & \theta\left(-0.0790151\kappa-0.297271-\frac{0.338563}{\kappa}-\frac{0.592782}{\kappa^{2}}-\frac{0.393406}{\kappa^{3}}\right.\nonumber \\
 & \left.-\frac{1.40262}{\kappa^{4}}+\frac{0.674301}{\kappa^{5}}-\frac{2.54331}{\kappa^{6}}-\frac{15.0062}{\kappa^{7}}+\frac{124.537}{\kappa^{8}}\right)\,.\label{eq:70a}\end{align}
For smaller $\alpha$, the convergence is even better, but this would
imply in a larger length scale of the modification, which may be not
natural. In Fig.\,\ref{fig:4} we present our results for fixed $\beta=2$
and different values of $\alpha$, showing that if $\alpha$ is taken
to be greater than one, the convergence of the $1/N$ expansion is
not adequate.

\section{\label{sec:Conclusions}Conclusions}

In this work, we studied the noncommutative quantum mechanics using
the $1/N$ expansion for the anharmonic and Coulombian potential.
We showed that, for a particular choice of the noncommutativity matrix
$\theta_{ij}$, we could apply the $1/N$ expansion for a noncommutative
potential depending only on a noncommutativity scalar parameter $\theta$.
With this simplification, we studied the anharmonic oscillator, calculating
the ground state energy up to the order $1/N^{12}$. For this potential,
the expansion presented good convergence properties.

For the Coulombian potential, however, the usual procedure of expanding
in powers of the noncommutative matrix is invalid leading to a divergent
$1/N$ expansion. In fact, the noncommutative modication to the potential,
\begin{equation}
\frac{Z\hat{e}^{2}}{2r^{3}}\theta m\,,\label{eq:71}\end{equation}
is highly singular near $r=0$. We therefore proposed a modified version
of the noncommutative Coulombian potential, where Eq.\,(\ref{eq:71})
is replaced by\begin{equation}
\left(1-e^{-\alpha r^{\beta}}\right)\frac{Z\hat{e}^{2}}{2r^{3}}\theta m\,.\end{equation}
The included term, $\frac{Z\hat{e}^{2}}{2r^{3}}e^{-\alpha r^{\beta}}\theta m$,
vanish exponentially for large $r$, so this modification is intended
to modify the potential in the $r\sim0$ region, softening the singularity
at the origin. We calculated the ground-state energy of such a modified
potential up to the order $1/N^{7}$, finding a good convergence for
certain values of $\alpha$ and $\beta$. The best choice for $\beta$
is $\beta=2$, and for $\alpha$, the range $0<\alpha<1$ provided
a well-behaved expansion.

We concluded that the $1/N$ expansion can indeed be applied in noncommutative
quantum mechanical systems, but it seems more sensitive to the singularities
of the potential than the usual perturbative expansion. 

\vspace{1cm}

\textbf{Acknowledgments. }This work was partially supported by the
Brazilian agencies Conselho Nacional de Desenvolvimento Cient\'{\i}fico
e Tecnol\'{o}gico (CNPq) and Funda\c{c}\~{a}o de Amparo \`{a} Pesquisa
do Estado de S\~{a}o Paulo (FAPESP). 

\pagebreak{}

\begin{thebibliography}{28}
\bibitem{foot1}Such a transcendental equation could only be solved
by numerical methods, but the method we are using is not well suited
for numerical calculations: as it becomes clear from Eqs.\,(\ref{eq:22})
to\,(\ref{eq:24}), at a given order $1/k^{n}$, the wavefunction
correction $U^{\left(n+1\right)}$ is a function possesing a potential
$0/0$ singularity at $u=1$, so it is difficult to evaluate numerically
such functions at $u=1$. This is a major stumbling block since the
energies at each order are calculated exactly by evaluating the r.h.s
of Eqs.\,(\ref{eq:22}) to\,(\ref{eq:24}) at $u=1$.
\end{thebibliography}
\end{document}